# Probing Plexciton Emission from 2D Materials on Gold Nanotrenches

Junze Zhou[1,†,*], P. A. D. Gonçalves[2,†], Fabrizio Riminucci[1], Scott Dhuey[1], Edward Barnard[1], Adam Schwartzberg[1], F. Javier García de Abajo[2,3,*], Alexander Weber-Bargioni[1,*]

[1]The Molecular Foundry, Lawrence Berkeley National Laboratory, 1 Cyclotron Road, Berkeley, CA 94720, USA

[2]ICFO – Institut de Ciències Fotòniques, The Barcelona Institute of Science and Technology, 08860 Castelldefels (Barcelona), Spain

[3]ICREA – Institució Catalana de Recerca i Estudis Avançats, Passeig Lluís Companys 23, 08010 Barcelona, Spain

[†] These authors contributed equally to the work

* Corresponding authors: junzezhou@lbl.gov, javier.garciadeabajo@nanophotonics.es, afweber-bargioni@lbl.gov

## Abstract

Probing strongly coupled quasiparticle excitations at their intrinsic length scales offers unique insights into their properties and facilitates the design of devices with novel functionalities. In this work, we investigate the formation and emission characteristics of plexcitons, arising from the interaction between surface plasmons in narrow gold nanotrenches and excitons in monolayer $WSe_2$. We study this strong plasmon–exciton coupling in both the far-field and the near-field. Specifically, we observe a Rabi splitting in the far-field reflection spectra of about 80 meV under ambient conditions, consistent with our theoretical modeling. Using a custom-designed near-field probe, we find that plexciton emission originates predominantly from the lower-frequency branch, which we can directly probe and map the local field distribution. We precisely determine the plexciton extension, similar to the trench width, with nanometric precision via collecting spectra at controlled probe locations. Our work opens exciting prospects for nanoscale mapping and engineering of plexcitons in complex nanostructures with potential applications in nanophotonic devices, optoelectronics, and quantum electrodynamics in nanoscale cavities.

## Introduction

The exploration of strongly coupled quasiparticle excitations in quantum materials has opened a new frontier in nanophotonics.[1–4] A prominent example of such excitations are plexcitons,[5,6] resulting from the interaction between plasmon resonances and excitons.[3,7,8] Plexcitons have been shown to produce large optical nonlinearities,[9] lead to ultrafast energy exchange,[10] and enable investigations of quantum phenomena at room temperature,[1,2,11] thus offering an enticing route to engineer light-matter interactions at the nanoscale. In recent years, the field has been fueled by the emergence of two-dimensional (2D) transition-metal dichalcogenides[8,12] (TMDs) as robust materials that host excitons with large binding energies and oscillator strengths,[7,13,14] making them exceptionally attractive for achieving strong coupling at room temperature and, thus, facilitating the exploration of plexcitonic phenomena.

Despite intense efforts to explore plasmon–exciton coupling in monolayer TMD/plasmonic hybrid systems, the nanoscale properties of plexcitons have so far been largely unexplored. The difficulty of accessing small plasmonic cavities and the limitations of conventional, diffraction-limited, photoluminescence (PL) setups have made probing these interactions extremely challenging.[15,16] Strong plasmon–exciton coupling has been primarily realized by placing plasmonic nanoparticles onto TMD-covered substrates[17–20] (including metallic ones, in the so-called nanoparticle-on-a-mirror setup[21]), making nanoscale characterization techniques based on scanning probes, such as scanning near-field optical microscopy or scanning tunneling microscopy, nearly impossible. Directly accessing and probing a plexciton in the near-field has remained a pending challenge, yet is key to advancing our fundamental understanding of plexciton physics and to the purposeful design of plexcitonic devices.

In this work, we present a combined far- and near-field investigation of plasmon–exciton coupling in WSe$_2$-covered gold nanotrenches. This platform enables direct access to the nanoscale properties of plexcitons and creates highly localized and easily tunable plasmon resonances that can be tailored to match the TMD exciton energy. Using near-field hyperspectral PL imaging, we unambiguously demonstrate the formation of plexcitons, their strong localization over individual nanotrenches, as well as their emission characteristics. Our findings are further supported by a comprehensive theoretical modeling of the complex nanoscale optical response of the plasmonic structure and the coupled plasmon–exciton complex. Our work not only overcomes previous limitations for directly mapping plexcitonic excitations at the nanoscale, but it also provides a unique platform for in-depth exploration of the properties and emission dynamics of plexcitons that could find applications in nanophotonic devices,[22] optoelectronics,[23] and quantum electrodynamics in nanoscale cavities.[24]

## Sample design and plexciton signature

The light–matter interaction is quantified through the coupling strength, which for a plexciton is $g = \boldsymbol{\mu}_e \cdot \mathbf{E}_p(\mathbf{r}_e)$, where $\boldsymbol{\mu}_e$ is the transition-dipole moment of the exciton and $\mathbf{E}_p(\mathbf{r}_e)$ the electric field of the plasmon at the exciton position. Therefore, plexciton formation requires a substantial spectral and spatial overlap between the plasmon and the exciton. The coupling is facilitated when the plasmon electric field aligns with the exciton dipole moment [i.e., parallel to the TMD plane for optically active (bright) excitons[25,26]].

To investigate plexciton formation while enabling near-field probe access, we have developed plasmonic cavities consisting of one-dimensional (1D) gold nanotrenches conformally coated with 2 nm of Al$_2$O$_3$ (to prevent quenching) together with a monolayer of WSe$_2$ placed on top (Fig. 1a). Such plasmonic cavities support resonances that provide large field enhancements at the opening of the nanotrench, and thus, near the 2D material (Fig. 1b). Moreover, the plasmon resonance features a strong in-plane dipole oriented across the trench ($x$-axis),[27–29] which appropriately aligns with the exciton's in-plane dipole moment. We fabricate our devices using well-established techniques (see Supp. Sec. S1),[27,30] guided by simulations and reflection measurements to precisely tune the plasmon resonance to the exciton wavelength (see Supp. Sec. S3). Under strong-coupling conditions, the plasmon and exciton hybridize forming two new plexciton states. We detect the signature of the plexciton through reflection measurements (Fig. 1d) in the far-field and via near-field photoluminescence (Fig. 1a–c,e).

In order to demonstrate that our platform provides enough strong coupling to create a plexciton, we start by performing polarization-dependent reflection measurements by illuminating the sample with a white light of linear polarization varying from 0° (along the $x$-axis, perpendicular to the trench) to 90° ($y$-axis, parallel to the trench). The results clearly show the presence of two well-resolved dips in the reflection data for 0° polarization, which progressively merge into a single dip at the bare exciton wavelength as the polarization is rotated to 90° (Fig. 1d). Such a behavior is a hallmark signature of strong plasmon–exciton coupling since the level splitting is only observed when the nanotrench plasmon is efficiently excited with near 0° polarization (see Supp. Fig. S18).

We then investigate the nanoscale optical response of plexcitons via scanning near-field optical microscopy (SNOM) using a fiber-based pyramidal probe fabricated with a nanoimprinting procedure.[31] We start by collecting the near-field using a bare dielectric probe to minimize the tip-induced perturbation introduced in the plasmonic resonance and prevent PL quenching, which are known issues when using metallic tips.[32–35] The probe allows us to characterize plexciton emission via hyperspectral PL imaging by exciting the sample with a continuous wave (CW) laser at 633 nm wavelength coupled through a glass fiber, where the light is tightly focused by the pyramidal probe at the end of the fiber into a nearly diffraction-limited spot.[31] The plexcitonic PL emission is recorded by raster-scanning the probe in the $xy$ plane, exhibiting enhanced emission at the nanotrench location indicated by the bright emission stripe in the inset image of Fig. 1e. This image shows the integrated PL intensity at each pixel (for the correlated height map, see Supp. Fig. S5c). The spectrum captured at the nanotrench (position 1 in the inset) displays a prominent peak at 763 nm (indicated by the vertical red-dotted line in Fig. 1e), which is consistent with the lower polariton (LP) absorption identified in Fig. 1d. We did not detect any observable PL emission from the upper polariton (UP) branch, likely because it is quenched by fast nonradiative relaxation, in agreement with previous strong-coupling studies.[12,19] When the scanning tip is moved away from the trench, the PL intensity decreases and its maximum blueshifts to ≈742 nm (see PL spectra recorded at position 2 and 3, 50 nm and 200 nm away from the trench, respectively), thus corresponding to ordinary PL emission from the bare exciton (see Supp. Fig. S1b). This behavior suggests that plexciton emission is strongly localized at the nanotrench as it is only detected when the probe is directly positioned over the trench, showing a peak wavelength that is consistent with the LP dip in the far-field reflection spectrum. We emphasize that LP plexciton emission is substantially stronger than emission associated with the surrounding uncoupled exciton. While the signal collected by the pyramidal probe comes from a region of approximately 300 nm in diameter,[31] the plexcitonic emission signal originates from a much smaller area near the nanotrench, as confirmed below. This strongly enhanced, localized near-field emission allows us to spectrally resolve the near-field PL from the plexciton with sub-diffraction spatial resolution.

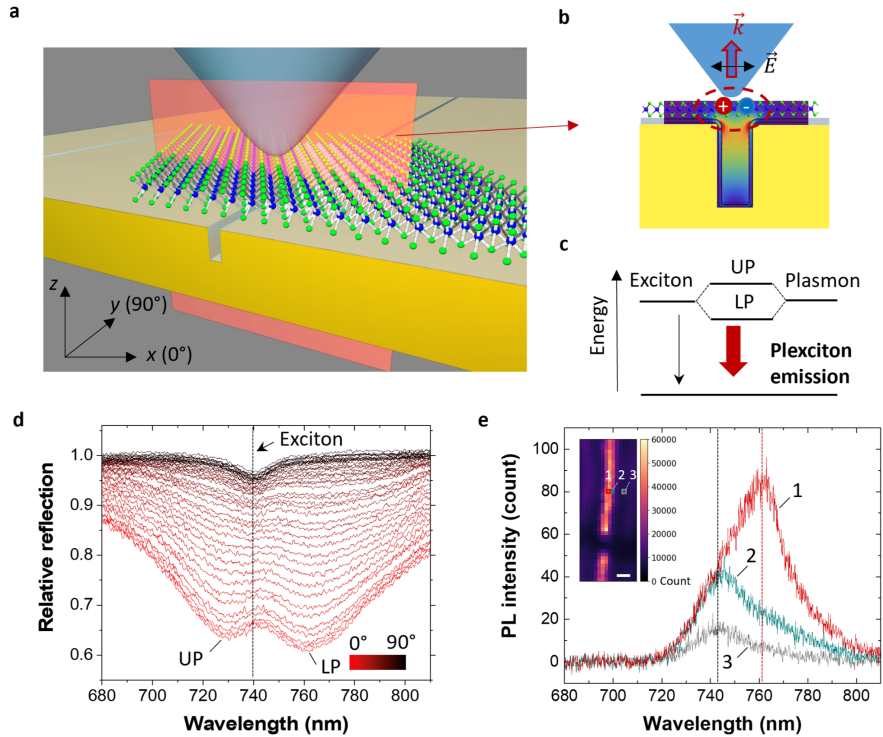

**Fig. 1 | Near-field probing and optical characterization of plexciton emission in a coupled $WSe_2$–Au-nanotrench system. a**, Schematic representation of the device, featuring a $WSe_2$ monolayer (not to scale) deposited on a gold nanotrench coated with an ≈2-nm-thick $Al_2O_3$ layer. The emitted light is collected using a sharp pyramidal probe (tip curvature radius of 20 nm) integrated into fiber facets (see Supp. Sec. S2 for additional details). **b**, Cross-sectional view (red plane in **a**) of the plasmon field intensity enhancement in the bare trench. **c**, Energy-level diagram illustrating the plexciton system, with the thick-red arrow denoting plexciton emission primarily originating from the lower polariton (LP), and the thin-dark arrow representing the emission of the bare (uncoupled) TMD exciton. **d**, Reflection spectrum as a function of incident-light polarization angle (0° to 90° for perpendicular-to-parallel to the trench). Each curve is normalized relative to measurements on unpatterned gold. **e**, Photoluminescence spectra acquired with our fiber-based pyramidal probe at positions 1-3 indicated in inset (scale bar is 200 nm).

## Tuning plasmonic resonances towards strong plasmon-exciton coupling

In order to analyze the plasmon–exciton coupling strength, we simulate the nanotrench plasmon resonance across the TMD exciton wavelength by tuning the trench width (see Supp. Sec. S3.C for details).[27,30,36,37] We then compare the anticrossing pattern imparted in the reflection spectra to these results. The resonance of a single nanotrench is controlled by its width and depth and features a Fabry–Perot-like resonance analogous to the lowest energy surface plasmon polariton (SPP) mode of a metal–insulator–metal (MIM) waveguide traveling along the trench depth,[37–39] corrected by the asymmetric truncation of the trench. In this picture, for a fixed nanotrench depth, narrower trenches are expected to exhibit plasmon resonances at larger wavelengths due to the stronger interaction between the SPPs at opposite trench walls. This intuition is supported by a comprehensive theoretical and experimental analysis of the dependence of the trench plasmon resonances on its geometry (width, depth, and shape) as well as the effect of the $Al_2O_3$ coating and PDMS residues (Supp. Sec. S4).

Before analyzing the coupled plasmon–exciton system, we first characterize the energy of the bare nanotrench plasmon and the uncoupled exciton of a $WSe_2$ monolayer on an unpatterned gold surface. The former is shown in Fig. 2a, where we identify a plasmon redshift in the reflection spectra corresponding to a decreasing nanotrench width, in good agreement with simulations (see Supp. Sec. S3). For example, the narrowest trench (≈19 nm wide) exhibits a clear reflection dip at $\lambda_p \approx 767$ nm; while for a width of 36 nm, the resonance wavelength moves to $\lambda_p \approx 642$ nm. This broad wavelength range allows us to tune the plasmonic resonance to sweep across the intrinsic exciton wavelength in the $WSe_2$ monolayer (742 nm, with a linewidth of ≈17 nm; see Fig. 1d at 90° polarization and also Supp. Fig. S1c).

Next, we transfer $WSe_2$ monolayers to the fabricated nanotrenches and study the formation of LP and UP branches through reflection measurements. Again, our measurements reveal a clear avoided crossing pattern in the reflection spectra (Fig. 2b–c). The symbols in Fig. 2c are the spectral minima of the reflection curves for two sets of measured samples (filled circles and stars) featuring nanotrenches with slightly different depths (see Supp. Fig. S13). From these data, we estimate a Rabi splitting of about 80 meV, which is comparable[17–20,40] or even surpasses[15,40] previous studies involving TMDs, including one in a similar configuration.[15] These observations are supported by simulations based on both the rigorous coupled-wave analysis (RCWA) technique[41–43] and the boundary-element method (BEM)[44] (see Supp. Sec. S5 for details). In particular, Fig. 2d shows the local density of optical states (LDOS) dispersion diagram calculated using BEM, to which we fit a coupled-oscillator model[1,7] to estimate the underlying plasmon–exciton interaction strength, yielding a Rabi splitting of 86.6 meV in excellent agreement with the experimental data (cf. Fig. 2c). In passing, we note that in our simulations we assumed 50-nm deep nanotrenches instead of the experimentally estimated 46 nm ± 2 nm to account for a systematic redshift observed in the measured data. We attribute this shift to the presence of PDMS residues remaining from the TMD transfer process (see Supp. Sec. S4.E).

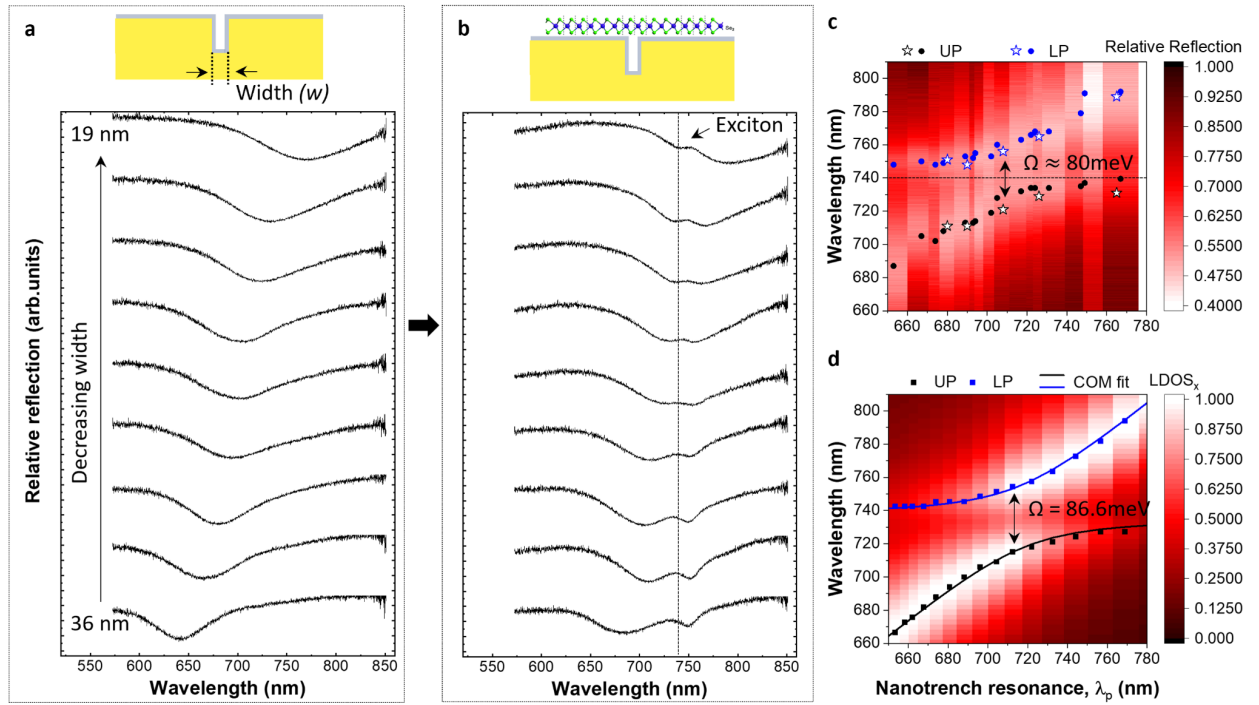

**Fig. 2 | Strong plasmon-exciton coupling**. **a**, Reflection spectra from bare gold nanotrenches (see top schematic) of varying width, normalized to flat gold. Curves are vertically offset for clarity. **b**, Reflection spectra of the $WSe_2$–Au-nanotrench coupled system, following

the same width order as in **a**. **c**, Wavelength dependence (horizontal axis) of the UP (black dots/stars) and LP (red dots/stars) as a function of plasmon resonance wavelength (vertical axis) identified in the uncovered nanotrench areas. The energy splitting at zero detuning is ≈80 meV. **d**, Calculated dispersion diagram of the LDOS based on the BEM, where LP and UP branches emerging from strong plasmon-exciton coupling are indicated by the blue and black squares. Fitting to a coupled oscillator model, renders a Rabi-like splitting of $\Omega \approx 80$ meV in experiment (**c**) and $\Omega \approx 86.6$ meV in theory.

## Near-field probing of plexciton emission

Building on the far-field characterization of plasmon–exciton coupling presented above, we now employ a fiber-based pyramidal tip to investigate the near-field optical response of the plexciton by hyperspectral mapping of its PL emission at the nanoscale. Figure 3a shows the PL emission collected by the tip while scanning laterally across the nanotrench at zero-detuning (i.e., when the plasmon resonance of the nanotrench matches that of the exciton in the TMD), revealing a pronounced near-field emission directly over the trench at a wavelength that matches the LP resonance observed in the far-field reflection spectrum. As in previous studies,[12,19] we do not detect any observable emission from the UP branch, presumably because it is quenched by nonradiative relaxation. To resolve the LP plexciton peak from the background, we fit the spectrum with two Lorentzians: one whose resonance and linewidth are fixed to those of the uncoupled emission peak (see Supp. Fig. S26) and another one with open fitting parameters. The result of this fitting procedure is presented in the inset of Fig. 3a, showing that plexciton emission is up to 5.75 times stronger than the background signal. This is notably stronger than reported in a previous study using scattering from a silver nanowire to readout the plexciton.[45] The fitted linewidth of the plexciton peak is slightly larger than that of the uncoupled exciton, which can be attributed to losses from the plasmon component in the plexciton.

Furthermore, we find that the measured near-field PL emission closely follows the reflection measurements of the lower plexciton branch as the plasmon resonance is detuned (Fig. 3b). This alignment is substantiated by comparing (i) a zoomed-in region of the calculated dispersion diagram (from Fig. 2d) with the LP dips in the measured reflection spectra superimposed (see also Supp. Fig. S25), and (ii) the corresponding near-field PL emission spectra. The latter are obtained by extracting and normalizing the plexciton spectra from the hyperspectral line scans (from Fig. 3a and Supp. Fig. S25). Notably, the plexciton peak positions from these samples are located at the same LP absorption peak wavelength as indicated by the arrows, which confirms the correlation between the near-field plexciton peak and the LP detected in far-field reflection.

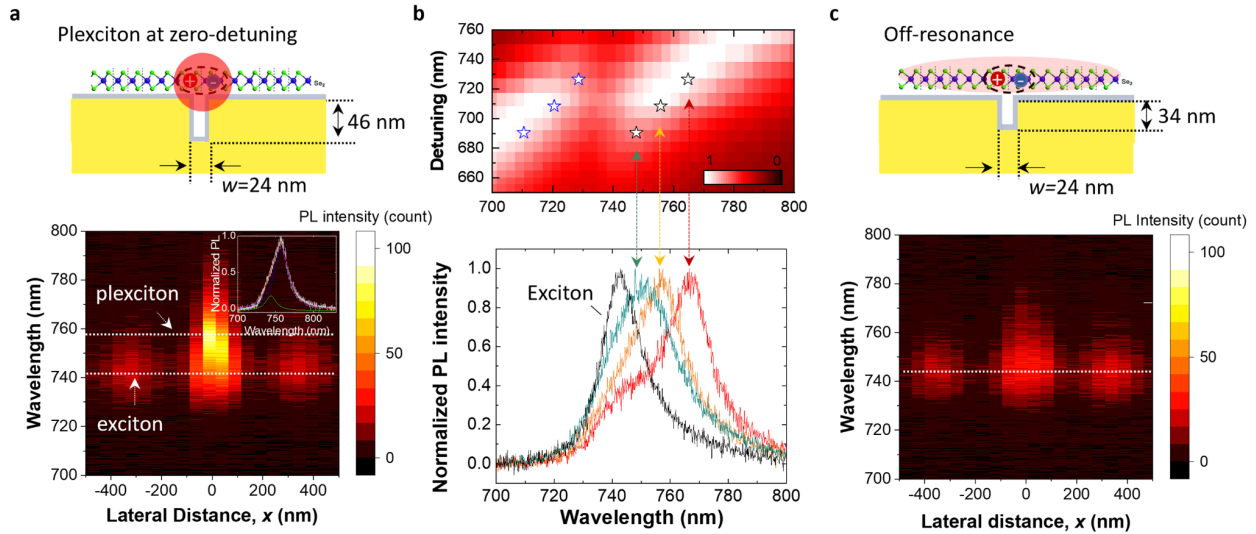

**Fig. 3 | Plexciton emission mapping and control experiment**. **a**, Schematic view of a plexcitonic sample with a trench width of 24 nm and depth of 46 nm, providing a zero-detuning plasmon-exciton coupling. The bottom plot is the corresponding line-scan analysis of PL spectra across the nanotrench, with the inset image showing the corresponding plexciton emission spectra deconvoluted with two peaks. **b**, Upper panel: segment of the simulated dispersion diagram presented in Fig. 2d (color plot), alongside the observed LP positions in reflection (marked as stars) for three nanotrenches of different widths (different detuning wavelength). Lower panel: the corresponding normalized

plexciton PL spectra, arranged from left to right, revealing uncoupled excitonic emission and strongly coupled exciton with plasmon resonances at different detuning wavelengths shown in the upper panel; line-scan analyses of the green and red spectra with detuning wavelength slightly deviating from the zero detuning, along with reflection spectra, can be found in Fig. S25. **c**, Schematic view of the system for a gold nanotrench whose plasmonic is off-resonance with the WSe$_2$ exciton (nanotrench with nominal $w = 24$ nm and $d = 34$ nm). The bottom plot is the corresponding line-scan analysis of PL spectra across the nanotrench.

While detuning the plasmon resonance alters the plexciton emission, other factors could also locally modify PL from the TMD layer, such as local strain[46] and screening.[47] To exclude these possibilities, we designed and measured a control sample with an off-resonance plasmon mode, maintaining the trench width at 24 nm and, thus, ensuring that the TMD layer experiences the same conditions over the trench, but reducing the trench depth to 34 nm, thereby setting the plasmon resonance to 655 nm (see also Supp. Sec. S6.A). Reassuringly, the corresponding hyperspectral linescan only shows PL emission at the bare, uncoupled exciton wavelength with no spectral evidence of the plexciton. Unlike in the coupled system, there is no significant PL enhancement over the nanotrench (Fig. 3a). Consequently, these observations rule out the effect of strain or probe-induced changes in the dielectric environment on the PL emission, further supporting our conclusions outlined above.

Having proven that the observed spectral signature is due to plexciton emission, we take advantage of this open-resonator platform combined with near-field characterization to directly probe, for the first time, the emission and field distribution of the plexciton. We leverage the capabilities of our near-field probe to investigate the spatial profile of the plexciton as well as the polarization state of its emission. Figure 4a illustrates the simulated electric field of the plexciton associated with the LP branch at zero detuning, showing that it is highly localized at the nanotrench and features a strong in-plane (across the trench) field component. Focusing on the area above the TMD (accessible by the near-field probe), we verify that the field is strongly confined both laterally and vertically. More quantitatively, we find a localization in $x$ with a FWHM of $\approx 20$ nm (same as the trench width) at $z = z_0$ ($z_0 \approx 10$ nm marks the initial scanning probe-sample distance[32]), and a vertical decay length within 20 nm of the TMD (see Fig. 4b and Supp. Fig. S23d). We then experimentally examine the spatial profile of the plexciton, shown schematically in Fig. 4c, by scanning the dielectric probe along $x$ for fixed $z = z_0$ and also along $z$ by retracting the probe away from $z_0$ (Figs. 4d and 4f, respectively). Figure 4d shows a zoomed-in hyperspectral line-scan (20 nm step-size) across the trench close to zero-detuning, featuring a strong emission enhancement centered around the trench and at a wavelength commensurate with the LP plexciton [linearly polarized across the trench (Fig. 4e), like the far-field data discussed above]. Incidentally, although the PL intensity is highest at the nanotrench's center, we still observe enhanced PL emission 100s of nm around the trench at the bare exciton wavelength, which is consistent with the optical resolution of the employed pyramidal probe.[31] We attribute this to weakly coupled (Purcell-type enhancement) excitonic emission. In contrast, the plexciton emission is exclusively observed when the probe is directly over the trench, with the PL peak shifting back to the bare exciton wavelength as the probe is moved as little as 20 nm away. Along the out-of-plane direction, PL emission from the plexciton could only be detected for probe distances within $\approx 10$ nm from the initial probe-sample distance and is thus localized within $\approx 20$ nm above the TMD (Fig. 4f).

The observed plexciton localization is in agreement with the simulated spatial distribution of the plexciton electric field (cf. Fig. 4b and Fig. 4d,f). These results not only reveal that the plexciton is highly confined at the trench position but also that its PL emission can only be efficiently detected using a near-field probe. The ability to control and manipulate the polarization of plexciton emission adds a degree of freedom for tailoring the optical properties of 2D materials, enabling novel functionalities and potentially enhancing device performance.

Lastly, to better resolve the plexciton from the weakly coupled emission, we employed a pyramidal probe coated with a 20-nm-thick gold layer (Fig. 4g). Such a probe provides a more confined excitation and collection profile due to the larger field enhancement at the probe apex (with a 20 nm radius of curvature).[32] This is demonstrated in Fig. 4h, which shows a hyperspectral line scan (15 nm step-size) exhibiting stronger localization of the enhanced PL emission at the plexciton wavelength owing to the improved resolution of the metal-coated probe. However, we also detected a small redshift of about 5 nm in wavelength when probing the plexciton emission using this probe. This indicates that the gold coating slightly modifies the plasmonic mode of the trench, which justifies our choice in favor of the dielectric probe for a cleaner characterization. Nevertheless, we still observe a localized plexciton PL emission, and the improved resolution of a metal-coated tip could be advantageous in some settings.

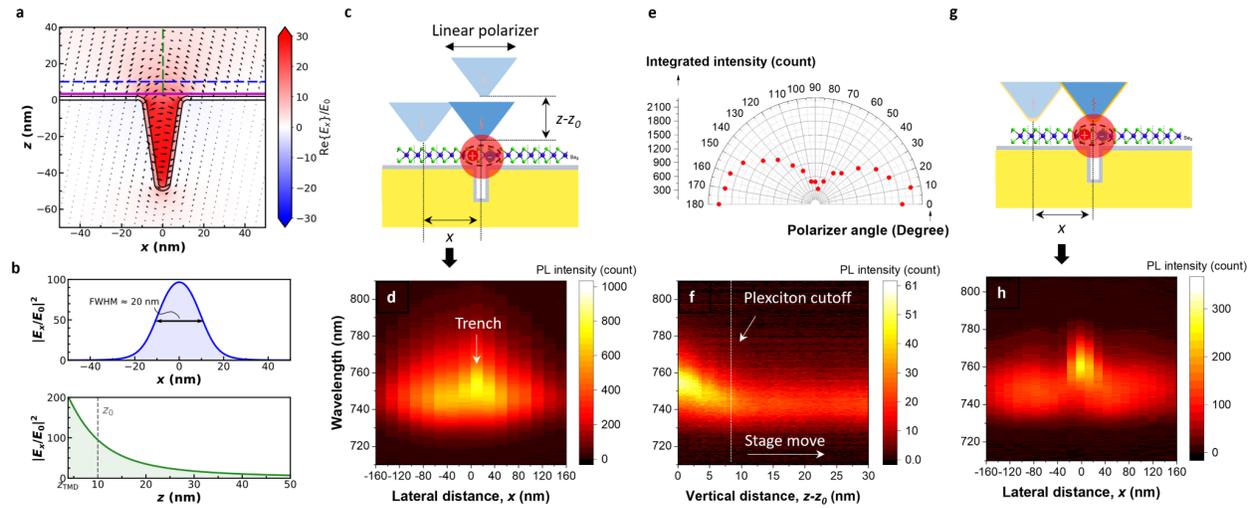

**Fig. 4 | Spatial localization and polarization properties of the plexciton**. **a**. Calculated electric field of the plexciton (LP) for a nanotrench corresponding to zero-detuning ($w = 20$ nm and $d = 50$ nm). The vector field is represented by the arrows (length proportional amplitude in log-scale). The background color plot shows the real-part of the $x$ component of the field. The ALD-coated nanotrench and the WSe$_2$ monolayer are outlined as black and purple lines. **b**. Corresponding field enhancement along the $x$ and $z$ directions (color-matching line cuts in **a**). **c**, Scheme depicting the scanning directions of the dielectric probe relative to the nanotrench site along the lateral ($x$) and vertical ($z$) directions. **d**, PL line scan along the lateral direction measured from a nanotrench at zero-detuning ($w = 24$ nm and $d = 46$ nm) with a step size of 20 nm. **e**, Polarization state of the plexciton assessed by rotating the linear polarizer at the output of the probe in increments of 10°. **f**, PL spectra recorded as the probe is retracted from the $z_0$ position with a step size of 0.5 nm. **g**, Similar to **c**, but now employing a gold-coated probe. **h**, PL map for a lateral line scan with a step size of 15 nm.

## Conclusion

In summary, we present a unique platform for investigating the nanoscale optical response of plexcitons by leveraging its capability for realizing strong plasmon–exciton coupling and consequent plexciton formation while also facilitating direct access to the plexciton near-field. In our plexcitonic device, the plasmon–exciton interaction is promoted by the large field enhancement with a strong in-plane component at the opening of a carved gold nanotrench provided by the trench's plasmon resonance, thus ensuring large spatial overlap to match the dipole-moment of the 2D TMD exciton. Far-field reflection measurements indicate a substantial energy splitting of 80 meV at ambient conditions due to plasmon-exciton hybridization. In addition, we perform a comprehensive near-field study using a scanning probe that allows us to directly investigate the source of the enhanced PL emission from the plexciton with a spatial resolution of a few tens of nanometers. We report enhanced near-field PL emission from the nanotrench site at wavelengths corresponding to the lower polariton resonance observed in the reflection spectra, supporting the conclusion that the detected emission originates from the lowest energy plexciton. This is further substantiated by measurements showing that the plexciton emission is strongly polarized across the trench direction. Our work opens new prospects for optical investigations of plexcitons and their emission dynamics at the nanoscale, creating a complementary technique to electron microscopy,[48] with the added advantages of lower cost, wider availability, and greater adaptability of optical setups which are particularly suited to examine the emission properties of polaritons. We believe our work could fuel further research into the near-field optical response of plexcitons in hybrid metal–TMD architectures, with potential applications in nanophotonic devices, optoelectronics, and quantum electrodynamics in nanoscale cavities.